\documentclass[preprint,showpacs,preprintnumbers,amsmath,amssymb]{revtex4}

\usepackage{graphicx}
\usepackage{dcolumn}
\usepackage{bm}
\usepackage{color}

\begin{document}

\title{DFT studies of ethylene in femtosecond laser pulses}

\author{Z.P. Wang,$^{1,2,3}$ P.M. Dinh,$^{3}$ P.-G. Reinhard,${^4}$ E. Suraud${^3}$ and F.S. Zhang$^{1,2}$\footnote{Corresponding
author. Email: fszhang@bnu.edu.cn}}
\address{$^1$The Key Laboratory of Beam Technology and Material
Modification of Ministry of Education, College of Nuclear Science
and Technology, Beijing Normal University,
Beijing 100875, People's Republic of China\\
$^2$Beijing Radiation Center, Beijing 100875, China\\
$^3$Laboratoire Physique Quantique (IRSAMC), Universit$\acute{e}$ P.Sabatier, 118 Route de Narbonne, 31062 Toulouse, cedex, France\\
$^4$Institut f$\ddot{u}$r Theoretische Physik,
Universit$\ddot{a}$t Erlangen, Staudtstrasse 7, D-91058 Erlangen,
Germany}
\date{\today }

\begin{abstract}
Using time-dependent density functional theory, applied to valence
electrons, coupled non-adiabatically to molecular dynamics of the
ions, we study the induced dynamics of ethylene subjected to the
laser field. We demonstrate the reliable quality of such an
approach in comparison to the experimental data on atomic and
molecular properties. The impact of ionic motion on the ionization
is discussed showing the importance of dealing with electronic and
ionic degrees of freedom simultaneously. We explore the various
excitation scenarios of ethylene as a function of the laser
parameters. We find that the Coulomb fragmentation depends
sensitively on the laser frequency. The high laser intensity can
cause brute-force Coulomb explosion and the laser pulse length
actually has influence on the excitation dynamics of ethylene.
\end{abstract}
 \maketitle

\section{introduction}

The developments in laser technology have opened the door to many
new research areas of physics, such as molecular physics. It is a
powerful tool to trigger and to analyze a bunch of dynamical
processes of molecules and clusters. These processes include
various dynamical situations, such as optical response \cite{U.
Kreibig}, multi-photon ionization \cite{Faisal}, and Coulomb
explosion of clusters \cite{J.Zweiback}. When the intensity of
laser is comparable to molecular electronic fields, the violent
excitations quickly drive the system beyond the Born-Oppenheimer
surface resulting some new processes such as high-order harmonic
generation \cite{J.Itatani, P.B.Corkum, P.Agostini}, bond
softening \cite{G.Yao, SK}, charge resonance enhanced ionization
\cite{T.Zuo, T.Seideman}. The theoretical understanding of these
mechanisms needs to solute the time-dependent Schr\"{o}dinger
equation for all electrons and all nuclear degrees of freedom.
However, this kind of numerical solutions only exist for smallest
systems, such as atoms \cite{JPHansen, Parker}, H$_2$ and H$_2^+$
\cite{Chelkowski, Kreibich, ADBandrauk, XBB, Liang}. For large
systems the computation is quite expensive due to the number of
degrees of freedom.

Recently, Horsfield \emph{et al.} use correlated electron-ion
dynamics to study the excitation of atomic motion by energetic
electrons \cite{Horsfield}. Saalmann \emph{et al.} have developed
a non-adiabatic quantum molecular dynamics (NA-QMD) to study
different non-adiabatic processes of different systems
\cite{Saalmann, T.Kunert, Kunert} and Calvo \emph{et al.} use a
combined methods to study the fragmentation of rare-gas clusters
\cite{F.Calvo}. In contrast, a fully-fledged coupled ionic and
electronic dynamics is developed. At the side of the electronic
dynamics, the well tested time dependent Density Functional Theory
at the level of the time-dependent local-density approximation
(TDLDA) \cite{NATO} is applied. The ions are treated by classical
molecular dynamics. This altogether provides a coupled TDLDA-MD,
for a review see \cite{F.Calvayrac}. Up to now, it has been
applied to study the dynamical scenarios of simple metal and
hydrogen clusters \cite{A. Castro, D. Dundas, E.Suraud, F Cal,
L.M.Ma, M.Ma, F.S.}. Besides these, organic molecules are
particularly interesting cases now, such as ethylene. For
ethylene, recently a lot of theoretical work has been carried out
to investigate either the optical response in linear domain
\cite{Takashi} or the dynamics of it without and with considering
the external laser field \cite{Ben-Nun, Kunert}.

In this paper, applying TDLDA-MD, we first devote to exploring the
impact of ionic degrees of freedom on the excitation dynamics and
we find the enhancement of the ionization with ionic motion.
Furthermore, we study the various excitation scenarios of ethylene
as a function of laser parameters. We find that these scenarios
depends sensitively on the laser parameters. It should also be
noted that switching from simple metals and hydrogen clusters to
organic molecules is a challenging test for the model.

The paper is outlined as follows. Section 2 provides a short
presentation of the theoretical and numerical approach. Section 3
first gives the ground-state properties of atoms and molecules.
Then the impact of ionic motion on the excitation dynamics is
discussed. Finally, the different excitation scenarios of ethylene
as a function of the laser parameters are presented. Section 4
gives conclusions.

\section{Theoretical and numerical methods}

The technicalities of the model we use have been presented in
detail elsewhere and for a review see \cite{F.Calvayrac}. Here for
the sake of completeness, we recall in this section the
ingredients and a few relevant formulas.

The molecule is described as a system of valence electrons and
ions. Valence electrons are treated by TDLDA, supplemented with an
average-density self-interaction correction (ADSIC)
\cite{C.Legrand}. Ions are treated as classical particles. They
interact via their repulsive Coulomb force. Pseudopotentials are
used for the interaction between ions and electrons. We apply the
form from \cite{Goedecker}. The TDLDA equations are solved
numerically on a grid 3D coordinate space (cuboid box size of
72*72*64) with a grid spacing of $\Delta =0.41$ a$_0$.

The computation of ground state wavefunctions is determined by the
damped gradient method \cite{F.Ca}. The time-splitting method is
applied to solve the time-dependent Khon-Sham equation for
electrons. For ions a Verlet algorithm is used for their
propagation under their mutual Coulomb interactions together with
the electron-ion forces as driven from the pseudo potential. For
the nonlocal part in the pseudopotentials, it is treated as a
third-order Taylor expansion of the exponential \cite{F.C}. The
absorbing boundary condition is employed to avoid periodic
re-feeding of the emitted electrons \cite{C.A.Ullrich}. For the
laser field, neglecting the magnetic field component, acting both
on electrons and ions, can be written as
\begin{eqnarray}
V_{laser} = E_{0}zf_{laser}(t)cos(\omega_{laser}t)
\end{eqnarray}
where $E_{0} \propto {\sqrt{I}}$, $I$ denoting the intensity, $z$
is the dipole operator, $f_{laser}(t)$ is the pulse profile and
$\omega_{laser}$ is the laser frequency. In this paper,
$f_{laser}(t)$ is chosen as $cosine^{2}$ in time. It should be
noted that this coupled ionic and electronic dynamics, constitutes
a true TDLDA-MD, and goes beyond the usual Born-Oppenheimer MD. It
is powerful to deal with the violent off-equilibrium dynamics as
induced by strong laser pulses. The capability of this technique
was demonstrate earlier calculations both on metal clusters
\cite{F Cal} and hydrogen clusters \cite{M.Ma}. For organic
molecules, in order to get the confidence of our approach to
describe them, we first start with the ground state calculation of
ethylene. Here we devote to the ionization potential (IP) of
Hydrogen, Carbon and ethylene. The IP which is important for
dynamical calculations is computed from the energy of the last
occupied electron state. In addition we calculate the bond lengths
of ethylene. For the more dynamical observables we study
excitation spectra of ethylene in the frequency domain which is
quite useful for exploring the excitation under the laser pulse.
The excitation spectra we get is quite similar to what Takashi
\emph{et al} \cite{Takashi} got. To avoid the repetition, in this
paper we give the location of the lowest peak and compare it with
the experimental data.

\section{Results and discussion}

\subsection{Ground state properties}
\begin{figure}[!ht]
\begin{center}
\includegraphics[width=8cm,angle=0]{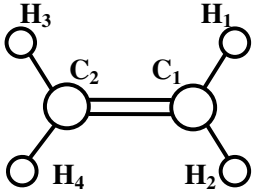}
\caption{Ionic structure of ethylene}
\end{center}
\end{figure}

The ethylene, C$_2$H$_4$, is the simplest organic $\pi$ system
holding $D_{2h}$ symmetry. There are 12 valence electrons. In our
calculation, it is placed in $x$-$y$ plane and the center of mass
is at the origin. The ionic structure of C$_2$H$_4$ is displayed
in Fig.~1.

\begin{table}
\caption{Ionization potentials (IP) for H, C and C$_2$H$_4$
obtained with LDA-SIC. In unit of eV. Experimental data are from
the website \cite{NIST}.}
\begin{tabular}{ccccccccccccccc}
\hline \hline
      &&H  &&  C  && C$_2$H$_4$ \\
      \hline
      exp [eV] && 13.6 && 11.3 && 11.0\\
      LDA-SIC [eV] && 14.0 && 11.3 && 11.5\\
      $\%$ error   && 2.9 && 0 && 4.5\\
\hline
\end{tabular}
\end{table}

\begin{table}
\caption{Theoretical bond lengths, in unit of a.u., and electronic
excitation energies of C$_2$H$_4$, in unit of eV, acquired from
LDA-SIC. Experimental data come from \cite{NIST, C.Petrongolo}.}
\begin{tabular}{ccccccccccccccc}
\hline \hline
      &&R$_{CC}$[a.u.]  &&  R$_{CH}$[a.u.]  &&  Lowest Peak [eV]\\
      \hline
      exp && 2.53 && 2.05 && 7.66\\
      LDA-SIC && 2.43 && 1.97 && 7.74\\
      $\%$ error   && 4.0 && 1.5 && 1.0\\
\hline
\end{tabular}
\end{table}

Table 1 shows the ionization potentials (IP) of neutral atoms and
C$_2$H$_4$. It is obvious that the IP are reproduced well when
compared with experimental values. Table 2 gives more detailed
results of ethylene. One can find that the relative errors are
less than 4$\%$ and the theoretical results are quite acceptable.

\subsection{Dynamics driven by laser pulse}

The laser field is coupled to ethylene as a time-dependent
classical field interaction simply via the dipole operator. As is
discussed in \cite{PG}, the laser pulse is characterized by
frequency, intensity and width which have their own specific
influence. The frequency determines the strength of the
laser-molecule coupling and the intensity determines the dynamical
regime. The pulse length determines the spectral resolution and
the interplay with ionic motion. Thus it is painful to choose the
test case in the enormous variety of conceivable laser parameters.
It should be noted that in all cases discussed following the laser
pulses are switched off at the moment of t=$T_{pulse}$.

\subsubsection{impact of ionic motion}

\begin{figure}[!ht]
\begin{center}
\includegraphics[width=8cm,angle=0]{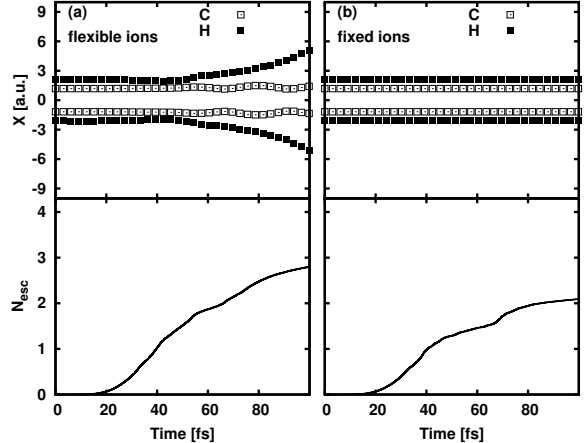}
\caption{A comparison of the excitation of ethylene induced by a
laser pulse with full ionic motion (a) and with fixed ionic
configuration (b). From the top to the bottom panels in both parts
are time evolution of ionic position along laser polarization and
the number of escaped electrons (carbon: hollow square, hydrogen:
full square). The laser polarization is along $x$ axis.
$I=10^{13}$~W/cm$^2$, $\omega_{laser}=8.16$~eV and
$T_{pulse}$=100~fs.}
\end{center}
\end{figure}

To investigate the effect of ionic motion on the excitation
dynamics, we compare calculations either with full ionic motion or
with fixed ionic configuration. Fig.~2(a) and (b) show the time
evolution of ionic motion and the number of escaped electrons
(N$_{esc}$) with full ionic and with fixed ionic motion
respectively. The difference between two cases is dramatic. From
Fig.2(a) we can see that ions start to move at around 40 fs. Four
hydrogen ions move more obviously than carbon ions. Electrons
start to escape gradually at around 18 fs. At around 50 fs, when
the laser reaches its highest amplitude, the electron emission
increases rapidly and there are almost 2.8 electrons emitted when
the laser is switched off. In Fig.2(b), when ions are kept fixed
during the calculation, the trend of electronic emission in the
first 40 fs is almost the same as in Fig.2(a). However, the
difference of N$_{esc}$ between two cases appears at around 40 fs,
which is the same moment at which ions start to move. Finally when
the laser is switched off, about two electrons are emitted. This
value is smaller than that of the moving ions case. The
enhancement of the ionization for the moving ions case can be
explained as in \cite{E.Suraud} even if the studied system here is
the covalent molecule. $\omega_{laser}$ is in the resonant region,
after increasing ionization, the optical spectrum has blue shift,
so $\omega_{laser}$ will be smaller than the resonant frequency.
However, the Coulomb force from increased net charge impels the
expansion of ions which could make red shift of the optical
spectrum and $\omega_{laser}$ becomes in the region of resonance
again.

The difference of electron emission above shows the importance of
treating electronic and ionic dynamics simultaneously. With the
coupling of electronic and ionic degrees of freedom we can explore
whatever excitation regimes.

\subsubsection{influence of laser frequency}

\begin{figure}[!ht]
\begin{center}
\includegraphics[width=8cm,angle=0]{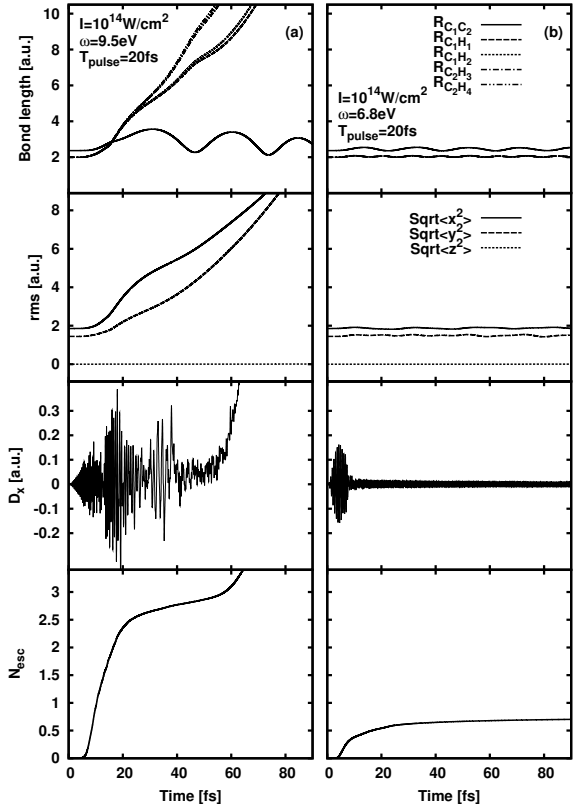}
\caption{Excitation of ethylene by two types of laser pulses with
different frequences. (a) $\omega_{laser}=9.5$~eV, (b)
$\omega_{laser}=6.8$~eV. $I=10^{14}$~W/cm$^2$ and
$T_{pulse}=20$~fs. The polarizations of lasers are both along the
$x$ axis. From the top to the bottom panels in both parts are the
time evolution of bond lengths, extensions of the ionic
distribution in three directions, dipole signal along the laser
polarization and the number of escaped electrons.}
\end{center}
\end{figure}

We now turn to the influence of the laser frequency on the
excitation dynamics of ethylene. We consider two frequency cases
here, one is $\omega_{laser}=9.5$~eV (Fig.~3(a)) which is in the
resonant region and the other one is $\omega_{laser}=6.8$~eV
(Fig.3(b)) which is off the resonant region. From the top to
bottom panels are the time evolution of the bond lengths, of the
ionic extensions along $x,y,z$ directions, which are
$\sqrt{\langle x^2 \rangle}$, $\sqrt{\langle y^2 \rangle}$,
$\sqrt{\langle z^2 \rangle}$, of the dipole signal along the laser
polarization, and of N$_{esc}$. The polarizations of laser pulses
are both along $x$ axis and $I=10^{14}$~W/cm$^2$. As the
$T_{pulse}=20$~fs is rather short, after laser is switched off, we
keep on following the tracks of ions and electrons until 100~fs.

From the top panel in Fig.~3(a), one can see that all of CH bond
lengths start to increase slowly in parallel at around 10 fs.
After the laser is switched off, they keep on increasing and start
to separate from each other at around 25~fs. All of them are
larger than 10~a.u. at around 60~fs. Such a large bond length
reflects that CH bonds are broken. The CC bond length (solid line)
starts to change a little later than that of R$_{CH}$ and
oscillates gently during the whole time without broken. For the
ionic expansion, as is shown in the second panel in Fig.~3(a), it
is obvious that ethylene expands within its plane and there is no
expansion in $z$ direction (dotted line). Compare the top two
panels in Fig.~3(b) with corresponding panels in Fig.~3(a), we can
find that for the off-resonant case, ions stay well around their
initial positions.

More physical insight can be gained by looking at the dipole
moment as shown in the third panel in both parts. When
$\omega_{laser}=9.5$~eV, it has no resemblance with the laser
pulse profile. The first obvious burst appears at around 10~fs.
This is almost the same moment at which the bond lengths start to
increase and the laser reaches its highest amplitude. At around
20~fs, when the laser is switched off, the dipole signal arrives
its highest amplitude and there are many bursts. It keeps on
oscillating strongly after the laser is switched off. At around 58
fs, a quick shift of the dipole signal takes place. This is due to
the fact that hydrogen ions are moving towards the border of the
calculation box. In contrast, for the off-resonant case, the
dipole signal has almost the same profile as the laser pulse and
it attains highest amplitude when the laser field reaches its
highest field. After the laser is switched off it remains
oscillating gently.

The dipole amplitude has consequence for the direct electron
emission, which is the same as in the cases of small metal
clusters \cite{M.Ma, C.A.}. A large dipole amplitude is related to
the strong emission while small amplitudes cause weak emission. As
we can see the electron emission in the bottom panels. In the
bottom panel in Fig.~3(a), the N$_{esc}$ starts to increase at
around 5 fs with a constant speed. This moment is a little earlier
than the movement of ions since electrons are more active than
ions. When the laser is switched off at 20~fs, there are almost
2.5 electrons escaped. The ionization continues but becomes much
more slowly after the laser is switched off. However, at around
58~fs, ionization increases quickly again. For the electron
emission in the bottom panel in Fig.~3(b), it starts to increase
at around 5~fs and levels off quickly after the laser is switched
off at 20~fs. The final value of N$_{esc}$ is much smaller than
that in Fig.~3(a).

From above discussion we can recognize that the laser frequency
plays an important role in the excitation dynamics of ethylene.
The frequency in the resonant region enhances the ionization
leading to the Coulomb fragmentation.

\subsubsection{influence of laser intensity}

\begin{figure}[!ht]
\begin{center}
\includegraphics[width=8cm,angle=0]{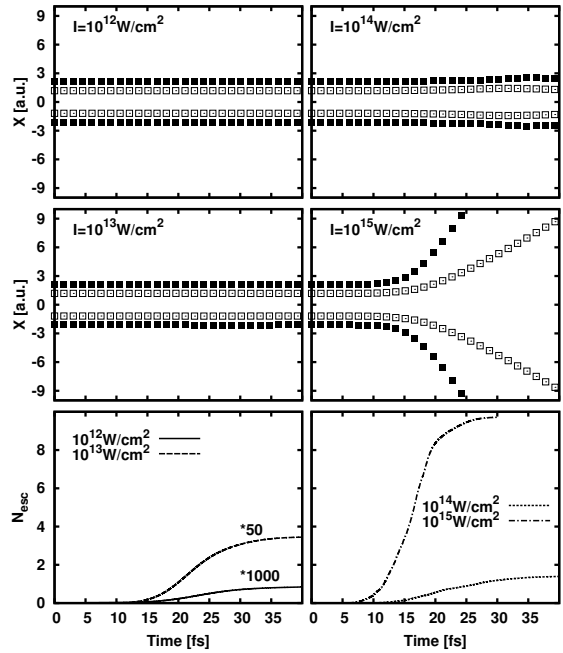}
\caption{The time evolution of ionic motion along the $x$
direction and ionization of ethylene with different laser
intensities. The laser polarization is along the $x$ direction,
$T_{pulse}=40$~fs and $\omega_{laser}=6.8$~eV.}
\end{center}
\end{figure}

We now keep the laser frequency and the pulse length fixed and
vary the laser intensity. Fig.~4 shows the excitation dynamics of
ethylene by the laser pulses with different intensities. The
$\omega_{laser}$ is 6.8~eV which is off-resonant and
$T_{pulse}=40$~fs. One can find that when intensities are low
(left panels), ions are almost frozen in the course of the laser
pulses and electron emissions in the bottom panel are quite small.
N$_{esc}$ values are multiplied by 1000 and 50 respectively in the
bottom panel in the left part in order to compare with N$_{esc}$
values in the bottom panel in the right part. In the right part,
for $I=10^{14}$~W/cm$^2$, ions seem to move a little at the end of
the laser pulse. The dissociation of ethylene occurs when
$I=10^{15}$~W/cm$^2$. Ions start to move at around 10~fs with a
constant velocity. It is clear that hydrogen ions move faster than
carbon ions. From the right bottom panel we can see that when
$I=10^{15}$~W/cm$^2$, eight to nine out of the twelve electrons
are emitted in a comparatively short time. The ethylene is grossly
Coulomb unstable at this charge state. As a consequence, the ionic
motion exhibits clearly the pattern of a straight Coulomb
explosion with CC double bond and CH bonds broken as shown in the
middle panel of the right part. Comparing the electron emissions
of different intensities, one can also find that for $I=10^{15}$
W/cm$^2$, the N$_{esc}$ (dash-dotted line) starts to increase at
around 7.5~fs. This moment is earlier than not only the ionic
motion but also that of other cases with lower laser intensities.

The results above show that the ionization of ethylene is enhanced
by the high laser intensity and the high intensity can cause
Coulomb explosion even if the laser frequency is in the
off-resonant region. Ions exhibit rigidity when laser intensities
are low.

\subsubsection{influence of pulse length of laser}

\begin{figure}[!ht]
\begin{center}
\includegraphics[width=8cm,angle=0]{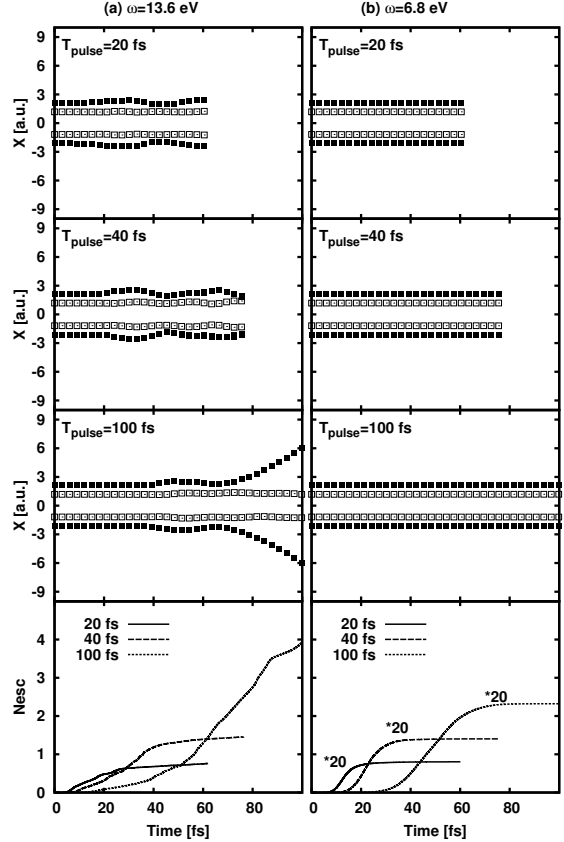}
\caption{Influence of laser pulse lengths on the excitation of
ethylene. $T_{pulse}=20$~fs, 40~fs, 100~fs. The laser
polarizations are along $x$ direction and $I=10^{13}$~W/cm$^2$.
The time evolution of ionic motion along the laser polarization
and ionization are presented. Two types of laser frequencies are
given. (a): $\omega_{laser}=13.6$~eV, (b):
$\omega_{laser}=6.8$~eV.}
\end{center}
\end{figure}

Christov \emph{et al} \cite{Christov} found that the harmonic
spectrum of argon taken with 25~fs laser pulses contains harmonics
up to 20 orders higher than for 100~fs laser pulses with the same
intensity due in part to the nonadiabatic response of the atomic
dipole to the fast rise time of the pulse. Here we explore the
influence of pulse lengths on the excitation of ethylene. Results
are shown in Fig.~5. We present two cases with different laser
frequencies, one is in-resonant, $\omega_{laser}=13.6$~eV
(Fig.~5(a)) and the other one is off-resonant,
$\omega_{laser}=6.8$~eV (Fig.~5(b)). We choose the moderate peak
intensity in both cases, $I=10^{13}$~W/cm$^2$. Since durations are
short for $T_{pulse}=20$~fs and 40~fs, after the laser is switched
off, we keep on tracking the dynamics of electrons and ions until
60~fs and 80~fs, respectively.

In Fig.~5(a) one can find that with the same intensity, when
frequency is in the resonant region, no matter how long the pulse
length is, ions move visibly and hydrogen ions move stronger than
carbon ions. The difference lies in that for short pulse lengths,
ions move gently in the course of laser pulses and the post-laser
ionic motion are also gentle. However, when $T_{pulse}=100$~fs,
after the first gentle oscillation, ethylene is dissociated with
CH bonds broken before the pulse is over. Therefore in this case,
the pulse length actually has strong influence on the ionic motion
and the long pulse length can cause fragmentation of ethylene. For
the electron emission, one can read off in the bottom panel that
electrons start to escape earlier than ionic motion and the
ionization is enhanced by the long pulse length. This is due to
the fact that with the long pulse length, active electrons absorb
energy from the laser field and multielectron excitation leads to
a highly excited ethylene with subsequently fragments, assisted by
Coulomb repulsion.

When $\omega_{laser}$ is in the off-resonant region, as shown in
Fig.~5(b), the pulse length has little effect on the ionic motion.
In top three panels, ions are rigid in the courses of the laser
pulses. The time evolutions of N$_{esc}$ are shown in the bottom
panel. In order to see curves clearly we multiply N$_{esc}$ values
by 20. It is clear that the ionization is enhanced by the long
pulse length. However, for different laser pulse lengths,
ionizations are quite small. Furthermore, one can find that the
electron emissions are saturated before the laser pulses are
switched off, no matter how long the pulse length is. The
independence of the excitation of ethylene on the laser pulse
length here is related to the fact that the laser frequency is
off-resonant and laser intensity is moderate, as a consequence,
the laser pulse is not strong enough to remove electrons and there
is no obvious ionic motion.

We can learn from the above discussion that when the laser pulse
is with moderate intensity and in-resonant frequency, the
excitation dynamics of ethylene depends sensitively on the pulse
length.

\section{Conclusions}

In this paper we have demonstrated the capability of TDLDA-MD to
describe the molecular dynamics and the ionic motion really has
strong effect on excitation dynamics. We have shown that this
approach produces reliable basic atomic and molecular properties
when compared with experimental values. We discussed various
excitation scenarios of ethylene subjected to different laser
pulses. These scenarios involve both electrons and ions but the
relative role of each species does depend on the actual excitation
conditions of the laser pulse. First, varying the laser frequency
with fixed laser intensity and pulse length indicates that the
appearance of deexcitation$/$explosion scenarios depends on the
relation of the laser frequency to the eigenfrequencies of the
system. Second, with fixed laser frequency and pulse length,
changing the laser intensity reveals that the high laser intensity
leads into the the brute-force Coulomb explosion of ethylene.
Finally, we find that with the same moderate intensity, the time
scales of electronic and ionic motions depend strongly on the
pulse length when laser frequency is in the resonant region. In
contrast, the excitation of ethylene is much less effected by the
pulse length when the laser frequency is off-resonant.

\section*{ACKNOWLEDGEMENTS}
This work was supported by the National Natural Science Foundation
of China (Grants No. 10575012 and No. 10435020), the National
Basic Research Program of China (Grant No. 2006CB806000), the
Doctoral Station Foundation of Ministry of Education of China
(Grant No. 200800270017), the scholarship program of China
Scholarship Council and the French-German exchange program PROCOPE
Grant No. 04670PG.

\end{document}